\documentclass[a4paper,11pt]{article}
\pdfoutput=1
\usepackage{jinstpub}
\usepackage{lineno}
\usepackage[alsoload=hep, per-mode=symbol, separate-uncertainty=true]{siunitx}
\usepackage{xspace}

\DeclareSIUnit[]\electron{\elementarycharge ^{-}}

\newcommand*{\GeV}{\ensuremath{\text{Ge\kern -0.1em V}}}
\DeclareSIUnit\clight{\text{\ensuremath{c}}}
\DeclareSIUnit[per-mode=symbol]\GeVc{\GeV\per\clight}

\title{Track reconstruction and matching between emulsion and silicon pixel detectors for the SHiP-charm experiment}
\collaboration{The SHiP Collaboration}
\emailAdd{nikolaus.owtscharenko@uni-siegen.de, markus.cristinziani@uni-siegen.de}
\date{November 2021}
\author[45]{C.~Ahdida,}
\author[49]{A.~Akmete,}
\author[15,d,h]{R.~Albanese,}
\author[7]{J.~Alt,}
\author[15,33,35,d]{A.~Alexandrov,}
\author[40]{A.~Anokhina,}
\author[19]{S.~Aoki,}
\author[45]{G.~Arduini,}
\author[39]{E.~Atkin,}
\author[30]{N.~Azorskiy,}

% ...
\affiliation[7]{ Physikalisches Institut, University of Freiburg, Freiburg, Germany }
% ...
\affiliation[15]{ Sezione INFN di Napoli, Napoli, Italy }
% ...
\affiliation[19]{ Kobe University, Kobe, Japan }
% ...
\affiliation[30]{ Joint Institute for Nuclear Research (JINR), Dubna, Russia }
% ...
\affiliation[33]{ P.N.~Lebedev Physical Institute (LPI RAS), Moscow, Russia }
% ...
\affiliation[35]{ National University of Science and Technology ``MISiS``, Moscow, Russia }
% ...
\affiliation[39]{ National Research Nuclear University (MEPhI), Moscow, Russia }
\affiliation[40]{ Skobeltsyn Institute of Nuclear Physics of Moscow State University (SINP MSU), Moscow, Russia }
% ...
\affiliation[45]{ European Organization for Nuclear Research (CERN), Geneva, Switzerland }
% ...
\affiliation[49]{ Middle East Technical University (METU), Ankara, Turkey }
% ...
\affiliation[d]{ Universit\`{a} di Napoli ``Federico II'', Napoli, Italy }
% ...
\affiliation[h]{ Consorzio CREATE, Napoli, Italy }
% ...

\abstract{%
In July 2018 an optimization run for the proposed charm cross section measurement for SHiP was performed at the CERN SPS. A heavy, moving target instrumented with nuclear emulsion films followed by a silicon pixel tracker was installed in front of the Goliath magnet at the H4 proton beam-line. Behind the magnet, scintillating-fibre, drift-tube and RPC detectors were placed. The purpose of this run was to validate the measurement's feasibility, to develop the required analysis tools and fine-tune the detector layout. In this paper, we present the track reconstruction in the pixel tracker and the track matching with the moving emulsion detector. The pixel detector performed as expected and it is shown that, after proper alignment, a vertex matching rate of \SI{87}{\percent} is achieved.}

\arxivnumber{2112.11754}

\begin{document}
\maketitle
\flushbottom

\section{Introduction}
\label{sec:intro}

Knowledge of the charm production cross section in a thick target is of key importance for the proposed SHiP~\cite{SHiPCollaboration2015} experiment. The prediction of charmed-hadron production is essential to establish the sensitivity to detect new particles and to make a precise estimate of the $\nu_\tau$ flux stemming from charm decays. Current information on charm production at a center-of-mass energy of $\sqrt{s} = \SI{27}{\GeV}$ is limited to measurements with thin targets~\cite{Lourenco2006}. For the determination of the flux of charmed hadrons the cascade production is of crucial importance and needs to be verified experimentally. The SHiP-charm project~\cite{Akmete2017} aims at measuring the double-differential cross section, $\text{d}^2\!\sigma/(\text{d}E\,\text{d}\theta)$, for charm production using the \SI{400}{\GeVc} primary proton beam, extracted from SPS to the H4 beam-line of the SPS North Area at CERN. The target consisted of a shorter replica of the SHiP SND detector, and is interleaved with emulsion cloud chambers (ECC) for tracking charm production and decays. This was followed by a magnetized tracking spectrometer and by a muon tagger. In July \num{2018}, an optimization run took place at the H4 beam-line. We address the challenge of reconstructing common tracks (and events) from the information recorded by the fundamentally different ECC and pixel detectors. This is complicated by the fact that the ECC detector carries no timing information and was moving relative to the beam and pixel in order to prevent overexposure during a given spill. In this paper, results of matching ECC tracks and vertices to downstream pixel tracks by means of a $\chi^2$ minimization of the residuals are presented.

\section{Experimental setup}

\begin{figure}[htbp]
\centering
\includegraphics[width=\textwidth]{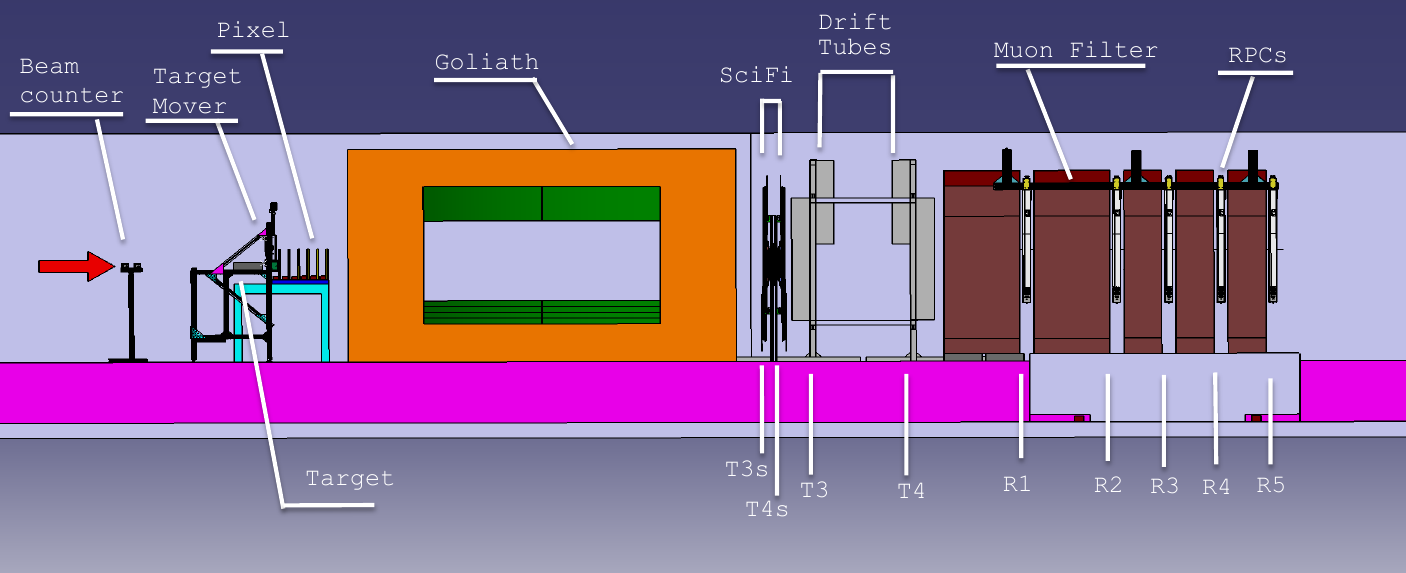}
\caption{Conceptual drawing of the SHiP-charm experiment setup for the 2018 test beam~\cite{charmReco}. The ECC is referred to as ``target''.\label{fig:setup}}
\end{figure}

The experiment was composed of three major parts: the ECC, the spectrometer and a muon tagger. For the measurement, \SI{400}{\GeVc} protons impinged on the ECC, made of tungsten sheets alternated with nuclear emulsion films. A detailed description of the ECC can be found in \cite{charmReco}. The most important properties are a very high spatial resolution and the permanence of each ionization trace. The permanent ionization makes it necessary to limit the occupancy in the emulsion films. The first electronic detector, \SI{1.8}{\centi\meter} downstream of the ECC, was the pixel detector. It was the first of three sub-detectors building the spectrometer together with GOLIATH~\cite{Goliath, Rosenthal:2310483}. Downstream of the magnet, a scintillating-fiber (SciFi) tracker of \SI{40 x 40}{\centi\meter} area per plane was positioned. It was followed by a drift-tube detector~\cite{Zimmermann:2006xr}, covering the outer regions of acceptance. The last detector was the muon tagger, built from resistive-plate chambers (RPCs) and an iron filter. Figure~\ref{fig:setup} displays the setup along the beam axis. Since the linking of analog ECC information with the pixel-detector tracks is crucial to the overall reconstruction and event selection, this paper focuses on this critical step. For the analysis described below, only the stand-alone data of these two detectors is used. While the ECC is passive, the pixel detector was triggered by the beam counter, a pair of scintillators requiring coincident detection of the primary beam protons.

\subsection{Beam conditions and data taking}

\begin{figure}[htbp]
    \centering
    \includegraphics[width=0.75\textwidth]{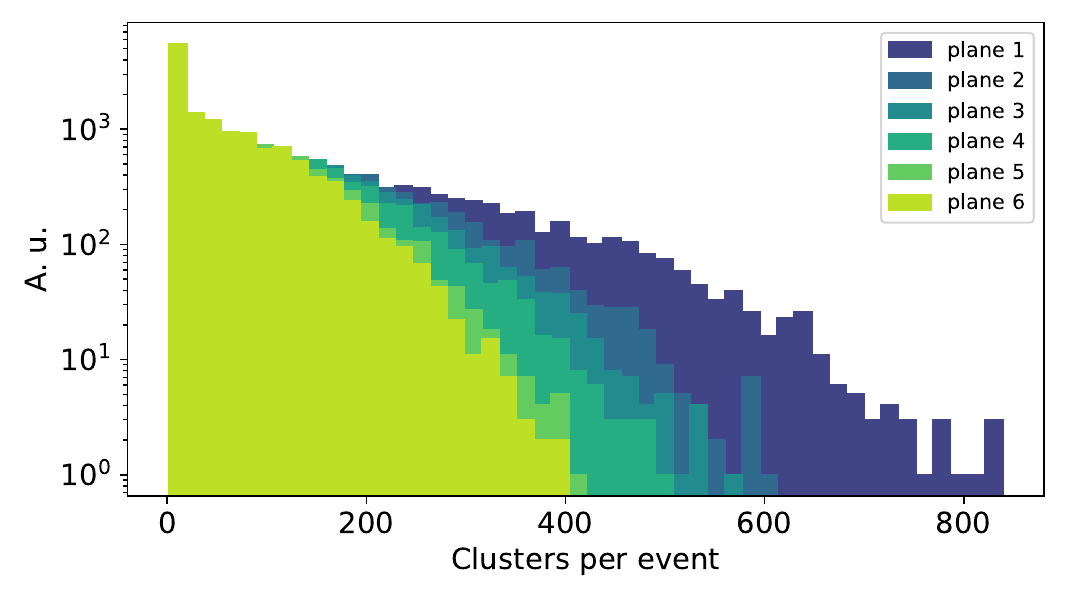}
    \caption{Overlay of histograms of the number of clusters per event in events with a proton interaction for all detector planes. The average for the most downstream plane 1 is \num{134} cluster per event with a maximum of \num{838}.}
    \label{fig:n_cluster}
\end{figure}

The beam in the North Area (and H4 beam-line) is slowly extracted in spills of mostly uniform duration of $\sim\!\SI{4.8}{\second}$. The beam was tuned to an elliptical shape with an extent of approximately \SI{2}{\centi\meter} in $y$ and \SI{0.7}{\centi\meter} in $x$\footnote{The coordinate system is defined such that the $z$-axis is parallel to the beam-line, the positive $y$-axis points upwards, while positive $x$ points to the right (direction ``Salève'') when looking downstream of the beam.
The most downstream emulsion layer is located at $z=0$.}. The number of protons per spill ranged from \num{7700} to \num{13800}. The occupancy limit on the ECC made it necessary to move the detector through the beam, and the beam shape was chosen to maximize the illuminated active area in this setup.
The pixel detector was synchronized via the analog start-of-spill signal, which was used to reset trigger counters and/or timestamps before each spill. The trigger counts all incoming protons and every trigger creates a new event.
Different target configurations were used for the SHiP-charm test-beam~\cite{charmReco}.
For this work one dedicated configuration is considered where the target consisted of \num{29} emulsion films interleaved with \num{28} tungsten sheets, adding up to a total passive-material budget of \SI{2.5}{\centi\meter} within the \SI{5}{\centi\meter} thick ECC. In this configuration the occupancy on the pixel detector for events with proton interaction was on average \num{86} cluster per plane per event, creating a high-occupancy environment for track reconstruction. Figure \ref{fig:n_cluster} shows the number of cluster per event for the single detector planes. During each of the five spills, the target moved at about \SI{\pm 2.6}{\centi\meter \per \second} along the horizontal axis, inverting the direction with every new spill. In between spills it was shifted upwards by \SI{2}{\centi\meter}, forming a snake-like pattern.

\subsection{Pixel detector}

As the most upstream element of the spectrometer, the pixel detector connects the analog information of the ECC with data taken by the other electronic sub-detectors: The ionization in the ECC is permanent, without any timing information. Tracks reconstructed in the pixel detector provide the necessary timestamp to fully reconstruct the event.

The pixel tracking detector consisted of six planes, each formed by two ATLAS IBL double-chip modules with hybrid pixels~\cite{Abbott2018}. ATLAS hybrid pixel detectors consist of a pixelated silicon sensor and the \mbox{FE-I4} read-out chip. The front-end chip offers an analog readout for every sensor pixel. They are electrically connected via solder bumps. Sensor and read-out chip were developed for a high-radiation and high-occupancy environment. The \mbox{FE-I4} clock runs at \SI{40}{\mega\hertz} which is therefore the maximum timing resolution.

One double-chip module is organized in \num{160} columns and \num{336} rows, resulting in \num{53760} pixels. The pixel pitch for the rows is \SI{50}{\micro\meter}, and \SI{250}{\micro\meter} for the columns. Pixels in the two central columns are \SI{450}{\micro\meter} wide to compensate for the small but necessary gap between the two independent front-end chips. To maximize the active area, edge columns are also wider, \SI{500}{\micro\meter}, with partially overlapping guard rings on the sensor. A double-chip sensor measures \SI{4.09 x 1.68}{\centi\meter}, resulting in an active area of about \SI{6.9}{\centi\meter\squared}. The sensors are about \SI{200}{\micro\meter} thick and were biased with \SI{-80}{\volt}. The sensor's hit detection efficiency is above \SI{99}{\percent}~\cite{ATLAS2012}. The front-end chips were tuned to a threshold corresponding to about \SI{1600}{\electron}.

The detector layout was optimized to achieve the best possible pointing resolution towards the ECC. The difference in pixel resolution between the $x$ and $y$ dimensions was compensated by rotating every other plane by \num{\pi/2} around the beam axis. This layout allows for three high resolution measurements in each dimension, $x$ and $y$, starting with \SI{50}{\micro\meter} resolution in $x$ on the first plane. Furthermore, the mounting and position of the planes relative to each other was optimized for maximal acceptance. To create a plane, two modules are placed on opposite sides of a single aluminium frame. This layout allows for a continuous active area. The aluminium frames were cut out to reduce the material budget as much as possible, while retaining the required mechanical stability and thermal contact.

\section{Track finding and reconstruction}
In the following we briefly discuss the independent track reconstruction in the ECC and the pixel detector, the alignment of the two detector systems with respect to each other and finally the  matching of common tracks.
\subsection{Emulsion detector}
Track reconstruction within the ECC is performed in two steps. First the emulsion films are scanned under a microscope to digitize the tracklets, second tracks are reconstructed from the tracklets with the \texttt{FEDRA} software~\cite{Tyukov:2006ny}. The intrinsic resolution of the emulsion films is \SI{0.7}{\micro\meter}~\cite{Arrabito:2007td} and the average film-by-film track efficiency was measured to be \SI{92(2)}{\percent}~\cite{charmReco}. The reconstructed tracks contain a set of at least two track segments, one for each emulsion plane. For track finding and fitting, a Kalman-Filter algorithm is used, taking into account inefficiencies in the reconstruction of track segments~\cite{Arrabito:2007td}. The track reconstruction purity was measured to be above \SI{95}{\percent}~\cite{Iuliano:2021}. Two-track vertices are identified with a criterion on the distance-of-closest-approach. They are associated to a common vertex based on a vertex probability taking into account the full covariance matrix of all involved tracks. Detailed information regarding the reconstruction is available in reference~\cite{charmReco}.

\subsection{Pixel detector}
Tracks in the pixel detector are reconstructed with a local pattern recognition starting from a track seed formed by two hits on the third and last detector plane. Track candidates are validated with a $\chi^2$ minimization fit. The pattern recognition only considers tracks with opening angles $\theta_{xz},\theta_{yz} \leq \SI{150}{\milli\rad}$, matching the spectrometer acceptance. A detailed description of the reconstruction can be found in \cite{emuPixMatching}. For the investigated run, \num{36132} events from \num{5} spills were recorded. The pixel detector efficiency is between \num{99.5} and \SI{99.9}{\percent}, while the tracking resolution is found to be \SI{15}{\micro\meter} in $x$ and \SI{26}{\micro \meter} in $y$ direction~\cite{emuPixMatching}.

\subsection{Alignment and track matching procedure}

In order to match track candidates between the pixel and the moving emulsion detectors, a set of good track candidates is selected and used
for a proper alignment. 
First, emulsion tracks are selected if they are associated to a vertex with at least six associated tracks. Tracks also have to feature a segment in the most downstream emulsion layer.
To minimize the influence of multiple scattering on the track resolution, only the track parameters of that most downstream segment are used in the following. In order to suppress tracks from fully penetrating protons (i.e., the beam), the number of track segments per track must be less than 29 (the total number of segments). 

The track parameters of interest for matching are the positions $x$, $y$ and the track angles
$\theta_{x}$ and $\theta_{y}$ of the furthest downstream track segment. The track angles $\theta_{x}$ and $\theta_{y}$ are required to be less than $\SI{150}{\milli \rad}$ each, in order only consider tracks within the spectrometer acceptance.
The time information provided with each reconstructed pixel detector track, $t= \mathrm{timestamp}\times \SI{25}{\nano\second}$, is used to translate the pixel detector's local coordinate system into the moving emulsion frame and transform the pixel track parameters $\boldsymbol{x}^{\mathrm{pix}} = (x, y, z, \theta_{x}, \theta_{y})$, accordingly. The uncertainty in the
time $t$ is $\mathcal{O}(\mathrm{ns})$, which is  small compared to the overall uncertainty coming from the spread in $x$ and the speed of the target mover $\mathcal{O}(\mathrm{ms})$, and is thus negligible. 

For the alignment, a set of eight parameters is introduced, $\boldsymbol{\alpha} = (x_0, y_0, z_0, \theta_{xz}, \theta_{yz}, \theta_{xy}, v_x, v_y)$, where $x_0, y_0, z_0$ are the offset of the pixel detector with respect to the emulsion reference frame, the two velocities $v_x$ and $v_y$ characterize the target mover, and the
rotations of the pixel detector about the $x$, $y$, and $z$ axes are denoted by $\theta_{yz}$, $\theta_{xz}$ and $\theta_{xy}$, respectively. The origin is set at the most downstream emulsion layer.
We define a track $\chi^2_{\mathrm{track}}$ of residuals between the emulsion and pixel detectors as 
\begin{equation}
    \chi^2_{\mathrm{track}} = \boldsymbol{r}^\mathrm{T} \boldsymbol{V}^{-1} \boldsymbol{r},
    \label{eq:chi2}
\end{equation}
where $\boldsymbol{r} = \boldsymbol{x}^{\mathrm{pix}} - \boldsymbol{x}^{\mathrm{ECC}} = (\Delta x, \Delta y, \Delta \theta_{x}, \Delta \theta_{y})$ is the vector of residuals and $\boldsymbol{V} = \boldsymbol{V}^{\mathrm{pix}} + \boldsymbol{V}^{\mathrm{ECC}}$ is the covariance matrix of residuals evaluated at the matching plane of $z=0$. The list of good track matches is constructed by calculating the $\chi^2$ of every possible pair between emulsion and pixel tracks. Only pairs with a $\chi^2<\num{100}$ are considered. Furthermore, there is a requirement on residuals of \SI{\pm 5}{\milli\meter} in $\Delta x$, and $\Delta y$ and a \SI{\pm 15}{\milli\rad} cut on the residuals in $\Delta \theta_{x}$, and $\Delta \theta_{y}$.  The list is then ordered in increasing values of $\chi^2$. A new list is created by starting from the beginning (smallest $\chi^2$) and moving down the list, at each step checking whether either the emulsion or pixel track were already used, in which case the pair would be removed from the list. This creates a set of good track matches with the minimal $\chi^2$ for a given set of alignment parameters. Whether this is the best possible list (minimal $\chi^2$) will depend on whether the two sub-detectors are properly aligned. The total 
\begin{equation}
    \chi^2 = \sum_{j} (\boldsymbol{r}^\mathrm{T} \boldsymbol{V}^{-1} \boldsymbol{r})_j,
     \label{eq:chi2_tot}
\end{equation}
is to be minimized, where the sum runs over track pairs $j$ between the emulsion and pixel tracks~\cite{BocciTRT}. The condition that the sample of tracks is minimal with respect to the alignment parameters can be written as
\begin{equation}
    0 \equiv \frac{\mathrm{d} \chi^2}{\mathrm{d}\boldsymbol{\alpha}} = 2 \sum_j \left(\frac{\partial \boldsymbol{r}^\mathrm{T}}{\partial \boldsymbol{\alpha}} \boldsymbol{V}^{-1}\boldsymbol{r} \right)_{j}.
\end{equation}
The optimal value of $\boldsymbol{\alpha}$ that satisfies this relation can be determined using the Newton-Raphson method. Given an initial set of alignment parameters $\boldsymbol{\alpha}_0$, an updated set $\boldsymbol{\alpha}_1$ is calculated as
\begin{equation}
    \boldsymbol{\alpha}_1 = \boldsymbol{\alpha}_0 - \left( \frac{\mathrm{d}^2 \chi^2}{\mathrm{d}\boldsymbol{\alpha}^2}\right)^{-1} \Biggr|_{\alpha_0} \left( \frac{\mathrm{d} \chi^2}{\mathrm{d}\boldsymbol{\alpha}} \right)\Biggr|_{\alpha_0}.
    \label{eq:alpha}
\end{equation}
This step is iterated until a convergence criterion is met, namely a minimal change in $\chi^2$ with increasing iterations. 
The alignment procedure can be summarized as follows:
\begin{enumerate}
\itemsep0em 
    \item Begin with an initial set of alignment parameters $\boldsymbol{\alpha}_0$.
    \item Calculate the $\chi^2$ per track pair with eq.~(\ref{eq:chi2}) and find the list of pairs with the minimal $\chi^2$.
    \item Calculate the total $\chi^2$ using eq.~(\ref{eq:chi2_tot}).
    \item Get a new set of alignment parameters $\boldsymbol{\alpha_1}$ using eq.~(\ref{eq:alpha}).
    \item Go back to Step 2 using $\boldsymbol{\alpha}_1$ in place of $\boldsymbol{\alpha}_0$ and repeat until the total $\chi^2$ converges.
\end{enumerate}

Convergence of the $\chi^2$ is not necessarily assured. If the misalignment is too large, the optimal set of track pairs could have a $\chi^2$ so large that it is dominated by combinatorial background, i.e., a pair with a large $\chi^2$ could take the place of an actual match if they share a track. Therefore, it is important to begin with a set of alignment parameters that are close to the optimal values. 

\section{Results}

\subsection{Track matching}
The alignment and matching procedure was performed on the data, where each spill was treated separately. 
Some alignment parameters are constrained by the initial mechanical alignment in the cavern before data taking, in particular the SHiP-charm setup was surveyed by the CERN EN/SMM group \cite{geometry_survey}.
The distance between the last emulsion layer and the first pixel layer was measured to be $z_0=\SI{1.8 \pm 0.1}{\centi\meter}$ while the speed of the target mover in the horizontal direction was measured to be $v_x = \SI[per-mode=symbol]{2.6 \pm 0.1}{\centi\meter\per\second}$. The value of $y_0$ changed depending on the spill and is estimated from the beam profile in $y$. The angles $\theta_{xz}$, $\theta_{yz}$ and $\theta_{xy}$ and the target mover speed in the vertical direction $v_y$ are initially set to 0. The alignment parameter $x_0$ is initially unknown, but can be estimated by setting $\Delta x = 0$ in the $\chi^2$ calculation and then looking for a peak in the resulting $\Delta x$ distribution after alignment.

\begin{figure}
    \centering
    \includegraphics[width=0.65\textwidth]{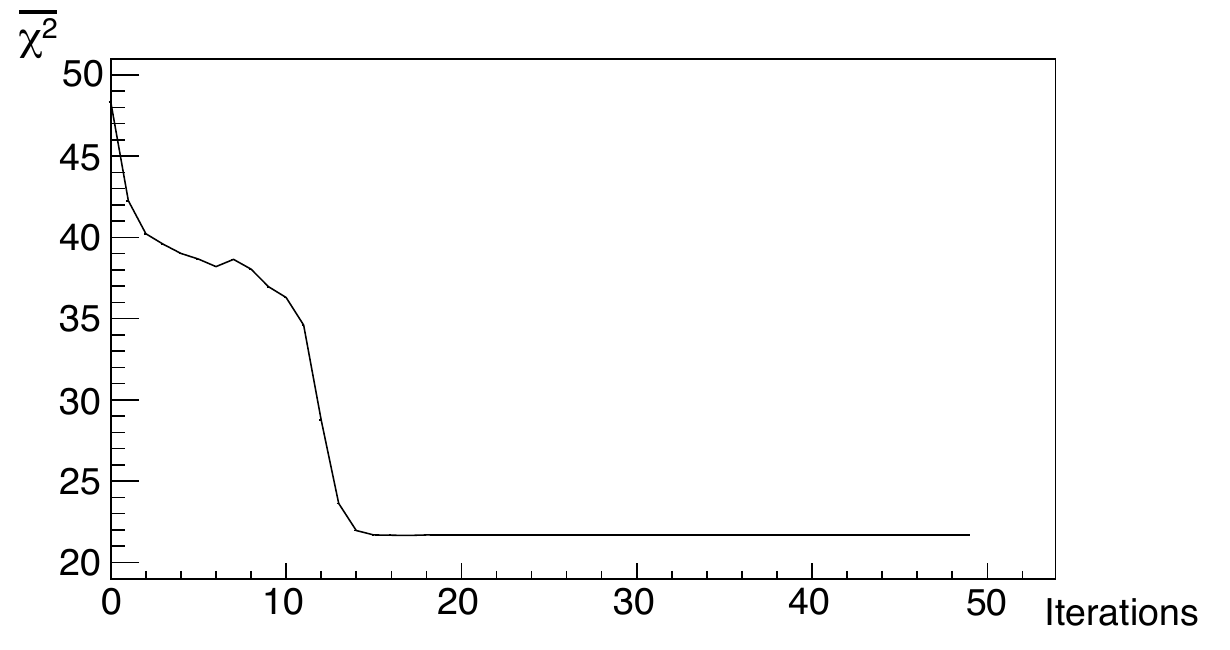}
    \caption{The average $\chi^2$ per track vs. the number of iterations of the alignment procedure.} 
    \label{fig:chi2}
\end{figure}

The evolution of the mean $\chi^2$ of all tracks is shown in Figure \ref{fig:chi2}, illustrating the improvement in the $\chi^2$ after alignment. 
The matching resolutions are found to be $\sigma_{x}$ = \SI{44}{\micro\meter}, $\sigma_{y}$ = \SI{80}{\micro\meter}, $\sigma_{\theta_{xz}} = \SI{4}{\milli\rad}$, and $\sigma_{\theta_{yz}} = \SI{3}{\milli\rad}$.

A shift in $x_0$ of about \SI{13.7}{\centi\meter} between alternate spills corresponds to the target moving $v_x\sim \SI[per-mode=symbol]{2.6}{\centi\meter\per\second}$ for \SI{5.2}{\second}, closely matching the target moving time, which included \SI{0.4}{\second} before/after the spill. 
Likewise, an observed shift in $y_0$ between spills can be explained by a vertical movement of $v_y\sim \SI[per-mode=symbol]{300}{\micro\meter\per\second}$. The distance between the last emulsion layer and first pixel layer $z_0$ is consistent with the survey measurement~\cite{geometry_survey}.
The angles $\theta_{xz}$ and $\theta_{yz}$ are close to 0, while $\theta_{xy}$ is about \SI{19}{\milli\rad}. Since $v_y$ changes sign between spills, this vertical velocity corresponds to a rotation of the emulsion brick with respect to the beam of about \SI{11}{\milli\rad}.     

\subsection{Physics performance} 
For the investigation of charmed hadronic interaction in SHiP-charm, a full event reconstruction including particle identification is necessary. This was achieved by measuring track deflection downstream of the magnetic field. Thus, the current analysis is focused on tracks which stay within the experiments acceptance and characteristic events are selected considering two main features.

\begin{figure}[h]
    \centering
    \includegraphics[width=0.49\textwidth]{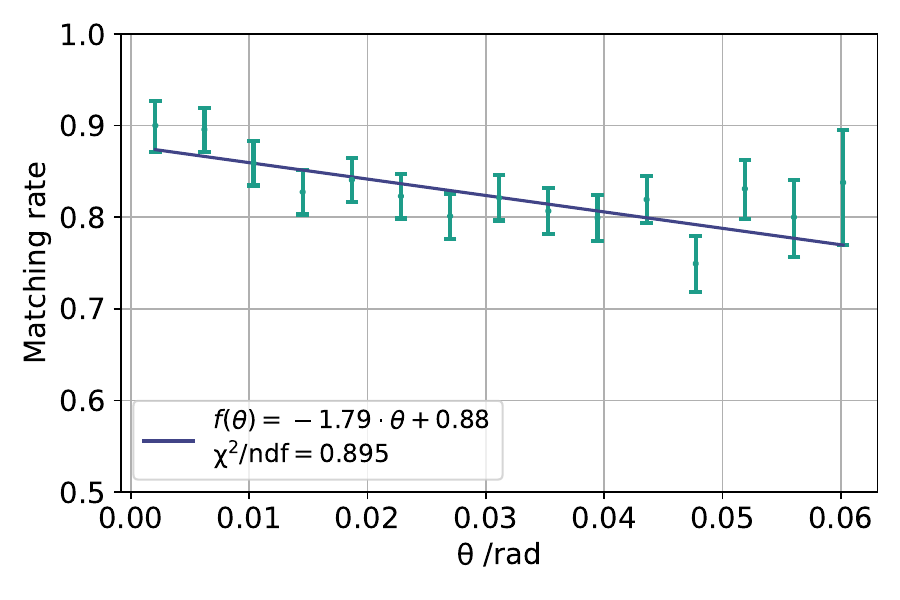}
    \includegraphics[width=0.49\textwidth]{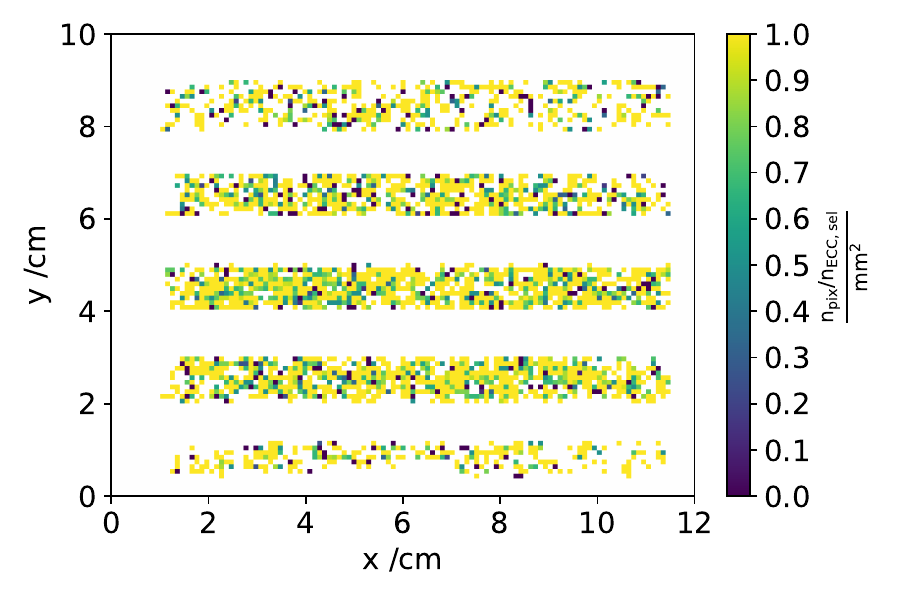}
    \caption{Track matching rate in the connected sub-detectors vs. track angle $\theta$ (left) and vs. $x$ and $y$ position of selected tracks (right). The non-uniform distribution of horizontal lines reflects the single spills.}
    \label{fig:match}
\end{figure}

In a first step only tracks from vertices with at least six tracks reconstructed in the ECC are selected. Secondly, the magnetic deflection of tracks beyond the SciFi detectors' acceptance is considered, and only tracks with opening angles smaller than \SI{62}{\milli\rad} are selected. The detector performance is then quantified in terms of the matching rate $\mathrm{\epsilon}$. Given a set of $n$ tracks, the matching rate is defined as the ratio of the number of ECC tracks matched in the pixel detector $n_{\mathrm{pix}}$ over the number of selected ECC tracks $n_{\mathrm{ECC, sel}}$:

\begin{equation*}
    \epsilon = \frac{n_{\mathrm{pix}}}{n_{\mathrm{ECC, sel}}}.
\end{equation*}

In Figure \ref{fig:match} the matching rate distributions for track matching after this selection are shown for the entire run. The average matching rate for selected emulsion tracks is \SI{82.6(4)}{\percent}. With the successful matching of at least one track a timestamp is assigned not only to the track but to the whole vertex. Thus, after matching, timestamps can be assigned to \SI{87}{\percent} of selected vertices. If a vertex is assigned a timestamp, the matching rate for tracks of this vertex is at least \SI{88}{\percent} on average, while for \SI{65}{\percent} of matched vertices all selected tracks are matched.

\begin{figure}
    \centering
    \includegraphics[width=0.75\textwidth]{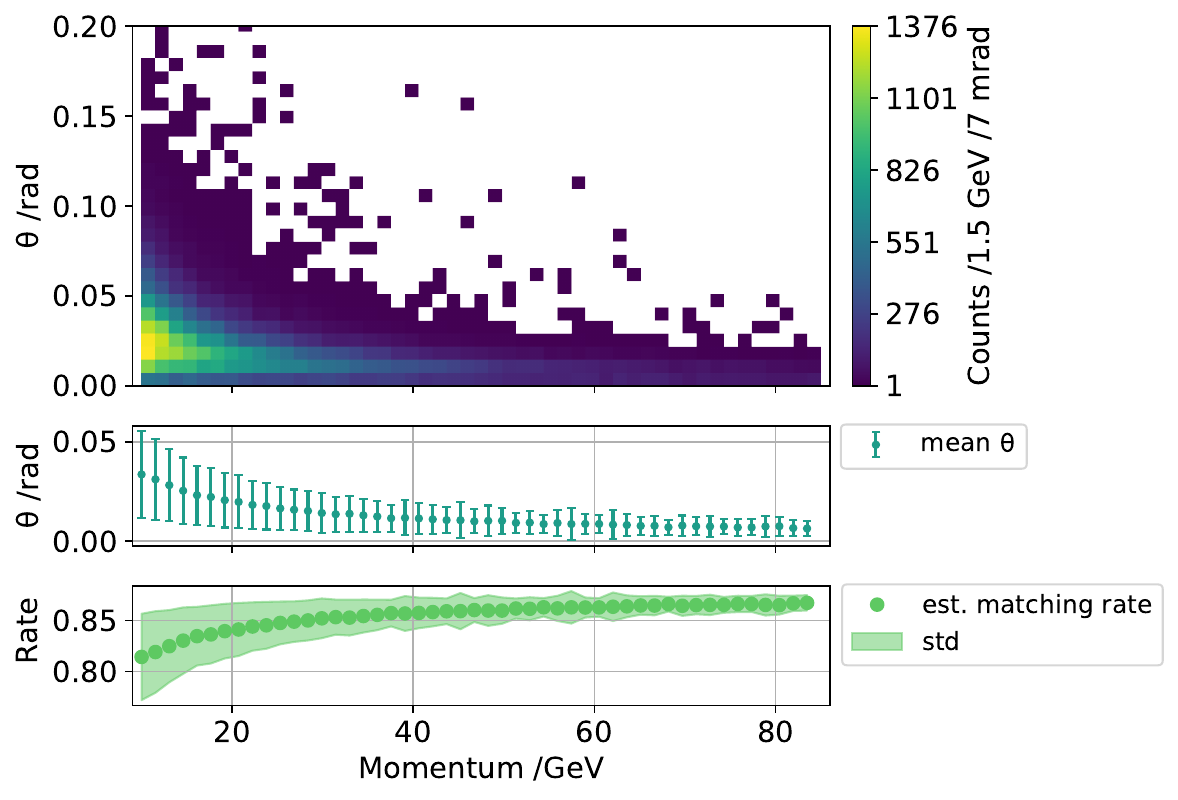}
    \caption{Track angle (top), average angle (center) and estimated matching rate (bottom) vs. momentum of Monte Carlo tracks in the emulsion. Only tracks  within the experiments acceptance are considered.}
    \label{fig:eff_estimate}
\end{figure}

The relation between track angle and matching rate can be used to estimate a matching rate with respect to the particles' momentum. In Figure~\ref{fig:match} (left) the rate is plotted for different track angles and a fit is performed. The uncertainties are computed by quadratic addition of the statistical uncertainty and the estimated systematic uncertainty.
The statistical uncertainty is computed as the 1-$\mathrm{\sigma}$ confidence interval of a binomial distribution, according to Bayes' theorem \cite{Casadei2012, Paterno2004}. The systematic uncertainty is determined using the mean difference of the matching rate for varying $\mathrm{\chi^2}$ constraints. The fitted model is applied to the average track angle for given momenta as obtained from a Monte-Carlo simulation of particle interactions in the ECC~\cite{charmReco}. Tracks with $p$ < \SI{10}{\GeV} are not considered, as these tracks are leaving the experimental setup after magnetic deflection.
The result is plotted in Figure~\ref{fig:eff_estimate}. For all tracks within the detector acceptance we expect a matching rate of at least \SI{81}{\percent}, increasing to \SI{87}{\percent} with track momentum. 

\section{Conclusion}
In this paper it is demonstrated that a moving emulsion detector without timing information and a stationary high-granularity pixel detector can successfully be used for track reconstruction in a high occupancy environment. The Newton-Raphson method is used to determine a set of eight alignment parameters. Two aspects were crucial for a successful alignment, a small distance between the two detectors and a set of adequate parameters to start the alignment procedure. With the described algorithm, \SI{82.6}{\percent} of the emulsion tracks within detector acceptance can be matched, corresponding to \SI{87}{\percent} of characteristic vertices. This proves the combination of ECC and pixel detector as well suited for a charm cross section measurement in the given setup. To evaluate whether the physics program can be met, a second optimization run and a study employing the whole spectrometer would be necessary. 

\acknowledgments

The SHiP Collaboration acknowledges support from the National Research Foundation of Korea, the Funda\c{c}\~{a}o para a Ci\^{e}ncia e a Tecnologia of Portugal, FCT; the Russian Foundation for Basic Research, RFBR and the TAEK of Turkey.
This work is supported by the German Science Foundation (DFG) through a research grant and a Heisenberg professorship under contracts CR-312/4-1 and CR-312/5-1.
We thank W.~Dietsche, F.~Hügging, J.~Janssen and D.~L.~Pohl of SILAB, Bonn for their assistance with the pixel detector.
The measurement reported in this paper would not have been possible
without a significant financial contribution from CERN. In addition, several member institutes
made large financial and in-kind contributions to the construction of the target and the
spectrometer sub-detectors, as well as providing expert manpower for commissioning, data
taking and analysis. This help is gratefully acknowledged.

\end{document}